\documentstyle[12pt,amssymbols]{article}
\input math_macros.tex

\begin{document}
\begin{titlepage}
\today          \hfill
\begin{center}
\hfill    LBL-36574 \\

\vskip .5in

{\large \bf Why Classical Mechanics Cannot Naturally Accommodate Consciousness
But Quantum Mechanics Can.}
\footnote{This work was supported by the Director, Office of Energy
Research, Office of High Energy and Nuclear Physics, Division of High
Energy Physics of the U.S. Department of Energy under Contract
DE-AC03-76SF00098.}
\vskip .50in

\vskip .25in
Henry P. Stapp

{\em Theoretical Physics Group\\
    Lawrence Berkeley Laboratory\\
      University of California\\
    Berkeley, California 94720}
\end{center}

\vskip .5in

\begin{abstract}
It is argued on the basis of certain mathematical characteristics that
classical mechanics is not constitutionally suited to accomodate
consciousness, whereas quantum mechanics is. These mathematical
characteristics pertain to the nature of the information
represented in the state of the brain, and the way this information enters
into the dynamics.
\end{abstract}
\vskip .3in
\centerline{\bf Prepared for a Special Issue of ``Psyche''}
\end{titlepage}
\renewcommand{\thepage}{\roman{page}}
\setcounter{page}{2}
\mbox{ }

\vskip 1in

\begin{center}
{\bf Disclaimer}
\end{center}

\vskip .2in

\begin{scriptsize}
\begin{quotation}
This document was prepared as an account of work sponsored by the United
States Government. While this document is believed to contain correct
 information, neither the United States Government nor any agency
thereof, nor The Regents of the University of California, nor any of their
employees, makes any warranty, express or implied, or assumes any legal
liability or responsibility for the accuracy, completeness, or usefulness
of any information, apparatus, product, or process disclosed, or represents
that its use would not infringe privately owned rights.  Reference herein
to any specific commercial products process, or service by its trade name,
trademark, manufacturer, or otherwise, does not necessarily constitute or
imply its endorsement, recommendation, or favoring by the United States
Government or any agency thereof, or The Regents of the University of
California.  The views and opinions of authors expressed herein do not
necessarily state or reflect those of the United States Government or any
agency thereof or The Regents of the University of California and shall
not be used for advertising or product endorsement purposes.
\end{quotation}
\end{scriptsize}

\vskip 2in

\begin{center}
\begin{small}
{\it Lawrence Berkeley Laboratory is an equal opportunity employer.}
\end{small}
\end{center}

\newpage
\renewcommand{\thepage}{\arabic{page}}
\setcounter{page}{1}

\noindent{\bf 1. Introduction}

\medskip

Classical mechanics arose from the banishment of consciousness from our
conception of the physical universe.
Hence it should not be surprising to find that the readmission of consciousness
requires going beyond that theory.

The exclusion of consciousness from the material
universe was a hallmark of science for over two centuries.
However, the shift, in the 1920's, from classical mechanics to quantum
mechanics marked a break
with that long tradition: it appeared that the only coherent way to
incorporate quantum phenomena into the existing science was to admit
also the human observer.$^{(1)}$
Although the orthodox approach of Bohr and the Copenhagen school was
epistemological rather than ontological, focussing upon ``our knowledge''
rather
than on any effort to introduce consciousness directly into the dynamics, other
thinkers such as John von Neumann$^{(2)}$, Norbert Weiner$^{(3)}$, and J.B.S.
Haldane$^{(4)}$ were quick to point out that the quantum mechanical aspects of
nature seemed tailor-made for bringing consciousness back into our conception
of matter.

This suggestion lay fallow for half a century. But the recent resurgence of
interest in the foundations of quantum theory has led
increasingly to a focus on the crux of the problem, namely the need to
understand the role of consciousness in the unfolding of physical reality.
It has become clear that the revolution in our conception of matter
wrought by quantum theory has completely altered the complexion of problem of
the relationship between mind and matter. Some aspects of this change were
discussed already in my recent book$^{(5)}$.
Here I intend to describe in more detail the basic differences between
classical mechanics and quantum mechanics in the context of the problem of
integrating consciousness into our scientific conception of matter, and to
argue that certain logical deficiencies in classical mechanics, as a foundation
for a coherent theory of the mind/brain, are overcome in a natural and
satisfactory way by replacing the classical conception of matter by a quantum
conception. Instead of reconciling the disparities between mind and matter by
replacing contemporary (folk) psychology by some yet-to-be-discovered future
psychology, as has been suggested by the Churchlands, it seems enough to
replace classical (folk) mechanics, which is known to be unable to account for
the basic physical and chemical process that underlie brain processes,
by quantum mechanics, which does adequately describe these processes.

\newpage
\noindent {\bf 2. Thoughts within the Classical Framework.}

\medskip

Thoughts are fleeting things, and our introspections concerning them are
certainly fallible.
Yet each one seems to have several components bound together by certain
relationships.
These components appear, on the basis of psycho-neurological data$^{(6)}$, to
be
associated with neurological activities occurring in different
locations in the brain.
Hence the question arises: How can neural activities in different locations in
the brain be components of a single psychological entity?

The fundamental principle in classical mechanics is that any physical system
can be decomposed into a collection of simple independent local elements each
of which interacts only with its immediate neighbors. To formalize this idea
let us consider a computer model of the brain. According to the ideas of
classical physics it should be possible to simulate brain processes by a
massive system of parallel computers, one for each point in a
fine grid of spacetime points that cover the brain over some period of time.
Each individual computer would compute and record the values of the components
of the electromagnetic and matter fields at the associated grid point.
Each of these computers
receives information only from the computers associated with neighboring
grid points in its nearly immediate past,
and forms the linear combinations of values that are the digital
analogs of, say, the first and second derivatives of various field values in
its neighborhood, and hence is able to calculate the values
corresponding to its own grid point.
The complete computation starts at an early time and moves progressively
forward in time.

On the basis of this computer model of the evolving brain I shall distinguish
the intrinsic description of this computer/brain from an extrinsic description
of it.

The intrinsic description consists of the collection of facts
represented by the aggregate of the numbers
in the various registers of this massive system of parallel computers: each
individual fact represented within the intrinsic description is specified
by the numbers in the registers in {\it one} of these computers, and the full
description is simply the conglomeration of these individual facts.
This intrinsic description corresponds to the fact that in classical mechanics
a complete description of any physical system is supposed to be specified by
giving
the values of the various fields (e.g., the electric field, the magnetic field,
 etc.) at each of the relevant spacetime points. Similarly, an intrinsic
description of the contents of a television screen might be specified by giving
the color and intensity values for each of the individual points (pixels)
on the screen, without any interpretive information (Its a picture of Winston
Churchill!), or any {\it explicit} representation of any relationship that
might exist
among elements of the intrinsic description (Pixel 1000 has the same values as
pixel 1256!). The analogous basic classical-physics description of a steam
engine would, similarly, give just the values of the basic fields at each of
the
relevant spacetime points, with no notice, or explicit representation, of the
fact that the system {\it can also be conceived of} as composed of various
{\it functional entities}, such as pistons and drive shafts etc.: the basic
or intrinsic description is the description of what the system {\it is},
in terms of its logically independent (according to classical mechanics)
local components, not the description of how it might be conceive of by an
interpreter, or how it might be described in terms of large functional entities
constructed out of the ontologically basic local components

I distinguish this intrinsic description from an extrinsic description.

An extrinsic description is a description that could be formed in the mind
of an external observer that is free to survey in unison, and act upon
together, all of the numbers
that constitute the intrinsic description, unfettered by the local rules of
operation and storage that limit the activities of the computer/brain. This
external observer is given not only the capacity to ``know'', separately, each
of the individual numbers in the intrinsic description; he is given also
the ability to know this collection of numbers as a whole, in the sense that
he can have a {\it single register} that specifies the entire collection of
numbers that constitutes the intrinsic description. The entire collection of
logically and ontologically independent elements that constitutes the intrinsic
description can be represented by a single basic entity in the extrinsic
description, and be part of the body of information that this external
observer can access directly, without the need for some compositional
process in the computer/brain to bring the information together from far-apart
locations. In general, {\it collections of independent entities} at
the level of the intrinsic description can become single entities
at the level of an extrinsic description.

The information that is stored in any {\it one} of the simple logically
independent computers, of which the computer/brain is the simple aggregate, is
supposed to be minimal: it is no more than what is needed to compute the local
evolution. This is the analog of the condition that holds in classical physics.
As the size of the regions into which one divides a physical system tends to
zero the dynamically effective information stored in each individual region
tends to something small, namely the values of a few fields and their first few
derivatives. And these few values are treated in a very simple way.
Thus if we take the regions of the computer simulation of the brain that
are represented by the individual local computers to be sufficiently small
then the information that resides in any {\it one} of these local computers
appears to be much less than information needed to specify a complex
thought, such as the perception of a visual scene: entries from many
logically independent (according to classical physics) computers must be
combined together to give the information contained in an individual thought,
which, however, is a single experiential entity. Thus the thought, considered
as a single whole entity, rather than as a collection of independent entities,
belongs to the extrinsic level of description, not to the intrinsic level of
description.

According to classical mechanics, the description of both the state of a
physical system and its dynamics can expressed at the intrinsic level.
But then how does one understand the occurrence of experientially whole
thoughts? How do extrinsic-level actual entities arise from a dynamics
that is completely reducible to an intrinsic-level description?

One possibility is that the intrinsic-level components of a thought are bound
together by some integrative process in the mind of a spirit being, i.e.,
in the mind of a ``ghost behind the machine'', of an homunculus. This
approach shifts the question to an entirely new realm: in place of the physical
brain, about which we know a great deal, and our thoughts, about which we have
some direct information, one has a new ``spirit realm'' about which science has
little to say. This approach takes us immediately outside the realm of science,
as we know it today.

Alternatively, there is the functional approach. The brain can probably be
conceived of, in some approximation, in terms of large-scale {\it functional
entities} that, from a certain global perspective, might seem to be controlling
the activity of this brain. However, in the framework of classical mechanics
such ``entities'' play no actual role in determining of the course of
action taken by the computer/brain: this course of action is completely
controlled by local entities and local effects. The apparent efficacy of the
large-scale ``functional entities'' is basically an illusion,
according to the precepts of classical mechanics, or the dynamics of the
computer/brain that simulates it: the dynamical evolution is completely fixed
by local considerations without any reference to such global entities.

As an example take a {\it belief}. Beliefs certainly influence, in some sense,
the activities of the human mind/brain. Hilary Putnam characterized the
approach of modern functionalism as the idea that, for example, a belief can
be regarded as an entry in a ``belief register'',
or a ``belief box'', that feeds control information into the computer program
that represents the brain process. Such a belief would presumably correspond,
physically, to correlations in brain activities that extend over a large part
of the brain. Thus it would be an example of a functional entity that a human
being might, as a short-hand, imagine to exist as a single whole entity, but
that, according to the precepts of classical mechanics, is completely
analyzable, fundamentally, into a simple aggregate of elementary and
ontologically independent local elements. The notion that such an
extrinsic-level
functional entity actually {\it is}, fundamentally, anything more than a
simple aggregate of logically independent local elements is contrary to the
precepts of classical mechanics. The grafting of such an actual entity onto
classical mechanics amounts to importing into the theory an appendage
that is unnecessary, nonefficacious, and fundamentally illusory from
the perspective of the dynamical workings of that theory itself.

Since this appendage is causally nonefficacious it has no signature,
or sign of existence, within classical physics. The sole reason for adding it
to
the theory is to account for our direct subjective awareness of it. Logically
and rationally it does not fit into the classical theory both because
it has no dynamical effects, beyond those due to its local components alone,
and because its existence and character contravenes
the locality principle that constitutes the foundation of the theory, namely
the principle that any physical system is to be conceived of as fundamentally a
conglomerate of simple microscopic elements each of which interacts only with
its immediate neighbors. Neither the character of the basic description of the
brain, within classical mechanics, nor the character of the classical
dynamical laws that supposedly govern the brain, provides any basis for
considering the brain correlate of a thought to be, at the fundamental as
distinguished from functional level, a single whole entity. One may, of course,
{\it postulate} some extra notion of ``emergence''. But nature must be able to
confer some kind of beingness beyond what is suggested by the precepts of
classical mechanics in order to elevate the brain correlate of a belief to the
status of an ontological whole.

This problem with `beliefs', and other thoughts, arises from the attempt to
understand the connection of thoughts to brains within the framework of
classical physics. This problem becomes radically transformed, however, once
one accepts that the brain is a physical system. For then, according to the
precepts of modern
physics, the brain must in principle be treated as a quantum system. The
classical concepts are known to be grossly inadequate at the fundamental level,
and this fundamental inadequacy of the classical concepts is not confined to
the molecular level: it certainly extends to large (e.g., brain-sized)
systems. Moreover, quantum theory cannot be coherently understood without
dealing in some detail with the problem of the relationship between
thoughtlike things and brainlike things: some sort of nontrivial
considerations involving our thoughts seems essential to a coherent
understanding of quantum theory.

In this respect quantum theory is wholly unlike classical physics, in which a
human consciousness is necessarily idealized as a non-participatory observer
--- as an entity that can know aspects of the brain  without
influencing it in any way. This restriction arises because classical physics
is dynamically complete in itself: it has no capacity to accomodate any
efficacious entities not already completely fixed and specified within its own
structure. In quantum theory the situation is more subtle because our
perceptions of physical systems are described in a classical language that is
unable to express, even in a gross or approximate way, the structural
complexity of physical systems, as they are represented within the theory:
there is a fundamental structural mismatch between the quantum mechanical
description of a physical system and our description of our perceptions of
that system. The existence of this structural mismatch is a basic feature of
quantum theory, and it opens up the interesting possibility of representing
the mind/brain, within {\it contemporary} physical theory, as a combination of
the thoughtlike and matterlike aspects of a neutral reality.

One could imagine modifying classical mechanics by appending to it the concept
of another kind of reality; a reality that would be thoughtlike,
in the sense of being an eventlike grasping of functional entities as wholes.
In order to preserve the laws of classical mechanics this added reality
could have no effect on the evolution of any physical system, and
hence would not be (publicly) observable. Because this new kind of reality
could have no physical consequences it could confer no evolutionary advantage,
and hence would have, within the scientific framework, no reason to exist.
This sort of addition to classical mechanics would convert it from a mechanics
with a monistic ontology to a mechanics with a dualistic ontology. Yet
this profound shift would have no roots at all in the classical mechanics onto
which it is grafted: it would be a completely {\it ad hoc} move from a
monistic mechanics to a dualistic one.

In view of this apparent logical need to move from monistic classical
mechanics to a dualistic generalization, in order to accomodate mind, it is a
striking fact that physicists have already established that classical mechanics
cannot adequately describe the physical and chemical processes that underlie
brain action: quantum mechanics is needed, and this newer theory, interpreted
realistically, in line with the ideas of Heisenberg, {\it already is
dualistic}. Moreover, the two aspects of this quantum mechanical reality accord
in a perfectly natural way with the matterlike and thoughtlike aspects of the
mind/brain. This realistic interpretation of quantum mechanics was introduced
by Heisenberg not to accomodate mind, but rather to {\it to keep mind out of
physics}; i.e., to provide a thoroughly objective account of what is happening
in nature, outside human beings, without referring to human observers and their
thoughts. Yet when this dualistic mechanics is applied to a human brain it can
account naturally for the thoughtlike and matterlike
aspects of the mind/brain system. The quantum mechanical description of the
state of the brain is automatically (see below) an extrinsic-level description,
which is the appropriate level for describing brain correlates of thoughts.
Moreover, thoughts can be identified with events that constitute
{\it efficacious choices}. They are integral parts of the quantum mechanical
process, rather than appendages introduced {\it ad hoc} to accomodate the
empirical fact that thoughts exist. These features are discussed in the
following sections.

\newpage
\noindent{\bf 3. Thoughts Within the Quantum Framework}

\medskip

Let us consider now how the brain would be simulated by a set of parallel
computers when the brain is treated as a quantum system.
To make this description clear to every reader, particularly those with
no familiarity with quantum theory, I shall start again from the classical
description, but spell it out in more detail by using some symbols
and numbers.

We introduced a grid of points in the brain.
Let these points be represented by a set of vectors $\vec{x}_i$, where $i$
ranges over the integers from 1 to $N$.
At each point $\vec{x}_i$ there was a set of fields $\psi_j(\vec{x}_i)$,
where $j$ ranges from 1 to $M$, and $M$ is relatively small, say ten.
For each of the allowed values of the pair $(i,j)$ the quantity
$\psi_j(\vec{x}_i)$ will
have (at each fixed time) some value taken from the set of integers that range
from $-L$ to $+L$, where $L$ is a very large number.
There is also a grid of temporal values $t_n$, with $n$ ranging from 1 to $T$.

The description of the classical system at any time $t_n$ is given, therefore,
by specifying for {\it each} value of $i$ in the set $\{1,2,...,N\}$
and {\it each} value of $j$ in the set $\{1,2,..., M\}$ {\it some } value of
$\psi_j(\vec{x}_i)$ in the set $\{-L, ..., +L\}$.
We would consequently need, in order to specify this classical system at one
time, $t_n, N\times M$ ``registers'' or ``boxes'', each of which is able to
hold
an integer in the range $\{-L, ..., +L\}$.

We now go over to the quantum mechanical description of this same system.
It is helpful to make the transition in two steps.
First we pass to the classical {\it statistical} description of the classical
system. This is done by assigning a probability to each of the possible states
of the classical system. The number of possible states of the classical system
(at one time) is $(2L+1)^{M\times N}$.
If the probability assigned to each of the possible classical systems is one of
$K$ possible values then the statistical description of the classical system at
one time requires $(2L+1)^{M\times N}$ registers, each with the capacity to
distinguish $K$ different values.
This can be compared to the number of registers that was needed to describe the
classical system at one time, which was $M\times N$ registers, each with a
capacity to distinguish $(2L +1)$ different values.

If the index $m$ runs over the $(2L+1)^{M\times N}$ possible classical systems
then a probability $P_m$ is assigned to each value of $m$,
where $P_m \geq 0$, and $\sum P_m =1$.

The quantum-mechanical description is now obtained by replacing each $P_m$ by
a complex number:
$$
P_m \Rightarrow r_m (\cos \theta_m + i \sin \theta_m),
$$
where $r_m = \sqrt{P_m},~\theta_m$ is an angle, $\cos \theta$ and $\sin\theta$
are the cosine and sine functions, and $i=\sqrt{-1}$.

This replacement might seem an odd thing to do, but one sees that this
description does somehow combine the particle-like aspect of things with a
wavelike aspect: the probability associated with any specific classical state
$m$ is $r^2_m =P_m$, and an increase of $\theta_m$ gives a wave-like
oscillation.

I am not trying to explain here how quantum theory works:  I am merely
describing the way in which the {\it description} of the computer/brain system
changes when one passes from the classical description of it to the quantum
description.

For the classical description we needed just $M\times N$ registers, but for the
quantum description we need $2\times (2 L +1)^{M\times N}$ registers.
Thus the information contained in the quantum mechanical description is
enormously larger. We need a value of $r_m$, and of $\theta_m$, for each of the
possible states of the {\it entire classical system}, where the specification
of the state of the classical system includes, simultaneously, a value of
$\psi_j(\vec{x}_i)$ for each allowed combination of values of $i$ and $j$.
That is, for each conceivable state of the {\it entire classical system} one
needs two separate registers.

Consider again a belief. As before, a belief would correspond physically to
some {\it combination} of values of the fields at many well-separated field
points $\vec{x}_i$. In the classical computer model of the brain there was no
register that represented, {\it or could represent}, such a combination of
values, and hence we were led to bring in an ``external knower'' to provide an
adequate ontological substrate for the existence of the belief.
But in the quantum-mechanical description there {\it is} such a register.
Indeed, each of the $2\times (2L+1)^{M\times N}$ registers in the quantum
mechanical description of the computer/brain corresponds to a possible
correlated state of activity of the {\it entire} classically-conceived
computer/brain.
Consequently, there is no longer any need to bring in an ``external
observer'': the quantum system itself has the requisite structural complexity.
Moreover, if we accept von Neumann's (and Wigners$^{(7)}$) suggestion that the
Heisenberg quantum jumps occur precisely at the high level of brain activity
that corresponds to conscious events then there is an ``actual happening'' (in
a
particular register, $m$) that corresponds to the occurrence of the conscious
experience of having an awareness of this belief. This ``happening'' is the
quantum jump that shifts the value of $r_m$ associated with this register $m$
from some value less than unity to the value unity. This jump constitutes the
Heisenberg ``actualization'' of the particular brain state that corresponds to
this belief. Jumps of this general kind are not introduced merely to
accommodate
the empirical fact that thoughts exist. Instead, they are already an essential
feature of the Heisenberg description of nature, which is the most
orthodox of the existing quantum mechanical descriptions of the physical world.
Thus in the quantum mechanical description of the brain no reference is needed
to any ``ghost behind the machine'': the quantum description already has
within itself a register that corresponds to the particular
state of the entire brain that corresponds to the belief. Moreover, it already
has a dynamical process for representing the ``occurrence'' of this belief.
This dynamical process, namely the occurrence of the quantum jump
(reduction of wave packet), associates the thought with a {\it choice}
between alternative classically describable possibilities, any one of which is
allowed to occur, according to the laws of quantum dynamics. Thus the dynamical
correlates of thoughts are natural parts of the quantum-mechanical
description of the brain, and they play a dynamically efficacious role in the
evolution of that physical system.

The essential point, here, is that the quantum description is automatically
wholistic, in the sense that its individual registers refer to states of the
{\it entire brain}, whereas the individual registers in the
classically conceived computer/brain represent only local entities. Moreover,
the quantum jump associated with the thought is a wholistic entity
: it actualizes as a unit {\it the state of the entire brain} that is
associated with the thought.

The fundamentally wholistic character of the quantum mechanical desription
nature is perhaps its most basic and pervasive feature.
It has been demonstrated to extend to the macroscopic (hundred centimeter)
scale in, for example, the experiments of Aspect, Grangier, and Roger$^{(8)}$.
In view of the fact that the wholistic character of our thoughts is so
antithetical to the principles of classical physics, it would seem imprudent to
ignore the wholistic aspect of matter that lies at the heart of contemporary
physics when trying to grapple with the problem of the connection of matter
to consciousness.
\newpage
\noindent {\bf 4. On The Thesis That `Mind Is Matter'.}

\medskip

Faced with the centuries-old problem of reconciling the thoughtlike
and matterlike aspects of nature many scientists and philosophers are turning
to the formula: `mind is matter'.$^{(9)}$ However, this solution has no content
until one specifies what matter is. The need to define `matter' is
highlighted by the extreme disparity in the conceptions of matter in classical
mechanics and quantum mechanics.

One might try to interpret the `matter' occurring
in this formula as the `matter' that occurs in classical physics. But this
kind of matter does not exist in nature.
Hence the thesis `mind is matter', with matter defined in this way, would
seem to entail that thoughts do not exist.

The thesis that `mind is matter' has been
attacked on the ground that matter is conceptually unsuited to
be identified with mind. The main rebuttal to this criticism given in ref. 9 is
that one does not know what the psychological theory of the future will be
like. Hence it is conceivable that the future theory of mind may not
involve the things such as `belief', `desire' and `awareness'
that we now associate with mind. Consequently, some {\it future} theory of
mind could conceivably allow us to understand how two such apparently disparate
things as mind and matter could be the same.

An alternative way to reconcile a theory of mind with the theory of
matter is not through some future conception of our mental life that differs
so profoundly from the present-day one, but rather through the introduction
the already existing modern theory of matter. Let me elaborate.

The main objection to the thesis that mind is matter --- as contrasted to the
view that mind and matter are different aspects of a single neutral
reality --- is based on the fact that each mind is known
to only one brain, whereas each brain is knowable to
many minds. These two aspects of the mind/brain
are different in kind: a mind consists of a sequence of private
happenings, whereas a brain consists of a persisting public structure. A
mind/brain has both a private inner aspect, mind, and a public outer aspect,
brain, and these two aspects have distinctive characteristics.

In the quantum description of nature proposed by Heisenberg reality
has, similarly, two different aspects. The first consists of a set of
`actual events': these events form a sequence of `happenings', each of which
actualizes one of the possibilities offered by the quantum dynamics .
The second consists of a set of `objective tendencies' for these events to
occur: these tendencies are represented as persisting structures in space and
time. If we correlate thoughts with high-level quantum events in brains, as
suggested by von Neumann, Wigner, and others, then we can construct a theory
that is a dual-aspect theory of the mind/brain, in the sense that
it correlates the inner, or mental, aspects of the mind/brain system with
`actual events' in Heisenberg's picture of nature, and it identifies the
the outer, or material, aspects of the mind/brain with the `objective
tendencies' of Heisenberg's picture of nature.

This theory might, on the other hand, equally well be construed as a theory in
which `mind is matter', if we accept the criteria for intertheoretic
reduction$^{(10)}$ proposed in
reference $9$. For this quantum theory of the brain is built directly upon the
concepts of the contemporary theory of matter, and it appears$^{(5)}$ to be
able to explain in terms of the laws of physics the causal connections
underlying human behavior that are usually explained in psychological terms.
Yet in this theory there is no abandonment of the normal psychological
conception of our mental life. It is rather the classical theory of matter
that is abandoned. In the terminology used in reference $9$ folk psychology is
retained, but folk physics is replaced by contemporary physics.
\newpage

\noindent {\bf 5. Final Remarks}

\medskip

It will be objected that the argument given above is too philosophical; that
the simple empirical fact of the matter is that brains are made out of neurons
and other cells that are well described by classical physics, and hence that
there is simply no need to bring in quantum mechanics.

The same argument could be made for electrical devices by an electrical
engineer, who could argue that wires and generators and antennae
etc. can be well described by classical physics. But this would strip him
of an adequate {\it theoretical} understanding of the properties
of the materials that he is dealing with: e.g., with a coherent and
adequate theory of the properties of transistors and conducting media, etc.
Of course, one can do a vast amount of electrical
engineering without paying any attention to its quantum theoretical
underpinnings. Yet the frontier developments in engineering today
lean heavily on our quantum theoretical understanding of the way electrons
behave in different sorts of environments.

In an even much more important way the processes that make brains work the way
they do depend upon the intricate physical and chemical properties of the
materials out of which they are made: brain processes depend in an exquisite
way on atomic and molecular processes that can
be adequately understood only through quantum theory.
Of course, it would seem easy to assert that small-scale processes
will be described quantum mechanically, and large-scale processes will be
described classically. But large-scale processes are built up in some sense
from small-scale processes, so there is a problem in showing how to reconcile
the large-scale classical behaviour with the small-scale quantum
behaviour. There's the rub! For quantum mechanics at the small scale simply
does not lead to classical mechanics at the large scale. That is exactly the
problem that has perplexed quantum physicists from the very beginning. One can
introduce, by hand, some arbitrary dividing line between small scale and large
scale, and decree that, in our preferred theory, the quantum laws will hold
for small things and the classical laws will hold for large things. But the
separation is completely ad hoc: there is no natural way to make this
division between small and large in the brain, which is a tight-knit physical
system of interacting levels, and there is no empirical evidence that supports
the notion that any such separation exists at any level below that at which
consciousness appears: all phenomena so far investigated can be understood by
assuming that quantum theory holds universally below the level where
consciousness enters.

Bohr resolved this problem of reconciling the quantum and classical aspect of
nature by exploiting the fact that the only thing that is known to be classical
is {\it our description of our perceptions of physical objects}. Von Neumann
and Wigner cast this key insight into dynamical form by proposing that the
quantum/classical divide be made not on the basis of size, but rather on the
basis of the qualitative differences in those aspects of nature that we call
mind and matter. The main thrust of ref. $5$ is to show, in greater detail, how
this idea can lead, on the basis of a completely quantum mechanical treatment
of our brains, to a satisfactory understanding of why our {\it perceptions} of
brains, and of all other physical objects, can be described in classical
terms, even though the brains with which these perceptions are associated are
described in completely quantum mechanical terms.. Any alternative
theoretical description of the mind/brain system that is consistent and
coherent must likewise provide a resolution to the basic theoretical problem of
reconciling the underlying quantum-mechanical character of our brains with the
classical character of our perceptions of them.

\newpage
\noindent {\bf 6. Conclusions}

\medskip

Classical mechanics and quantum mechanics, considered as conceivable
descriptions of nature, are structurally very different. According to
classical mechanics, the world is to be conceived of as a simple aggregate
of logically independent local entities, each of which interacts only with its
very close neighbors. By virtue of these interactions large objects and systems
can be formed, and we can identify various `functional entities' such as
pistons
and drive shafts, and vortices and waves. But the precepts of classical physics
tell us that whereas these functional units can be identified by us, and can be
helpful in our attempts to comprehend the behaviour of systems, these units do
not thereby acquire any special or added ontological character: they continue
to be simple aggregates of local entities. No extra quality of beingness is
appended to them by virtue of the fact that they have some special functional
quality in some context, or by virtue of the fact that they define a spacetime
region in which certain quantities such as `energy density' are greater than in
surrounding regions. All such `functional entities' are, according to the
principles of classical physics, to be regarded as simply consequences of
particular configurations of the local entities: their functional
properties are just `consequences' of the local dynamics; functional properties
do not generate, or cause to come into existence, any extra quality or kind of
beingness not inherent in the concept of a simple aggregate of logically
independent local entities. There is no extra quality of `beingness as a whole'
, or `coming into beingness as a whole' within the framework of classical
physics. There is, therefore, no place within the conceptual
framework provided by classical physics for the idea that certain patterns
of neuronal activity that cover large parts of the brain, and that have
important functional properties, have any special or added quality of
beingness that goes beyond their beingness as a simple aggregate of local
entities. Yet an experienced thought is experienced as a whole thing. From the
point of view of classical physics this requires either some `knower' that is
not part of what is described within classical physics, but that can `know' as
one thing that which is represented within classical physics as a simple
aggregation of simple local entities; or it requires some addition to the
theory that would confer upon certain functional entities some new quality not
specified or represented within classical mechanics. This new quality would be
a quality whereby an aggregate of simple independent local entities that
{\it acts} as a whole (functional) entity, by virtue of the various local
interactions described in the theory, {\it becomes} a whole (experiential)
entity. There is nothing within classical physics that provides for two such
levels or qualities of existence or beingness, one pertaining to persisting
local entities that evolve according to local mathematical laws, and one
pertaining to sudden comings-into-beingness, at a different level or quality
of existence, of entities that are bonded wholes whose components are
the local entities of the lower-level reality. Yet this is exactly what is
provided by quantum mechanics, which thereby provides a logical framework that
is perfectly suited to describe the two intertwined aspects of the mind/brain
system.
\newpage
 \noindent{\bf{Appendix A.
Salient Features of the Quantum Theory of the Mind/Brain Described in Ref. 5.
}}

1. {\underline{Facilitation}}.  The  excitation of a  pattern of neural firings
   produces   changes in  the  neurons that  have  the effect  of  facilitating
   subsequent excitations the pattern.

2. {\underline{Associative Recall.}} The facilitations mentioned above have the
   feature that the excitation of a part  of the pattern tends to spread to the
   whole pattern.

3. {\underline{Body-World   Schema.}} The  physical body of  the person and the
   surrounding  world  are  represented by  patterns of  neural  firings in the
   brain: these patterns  contain the information  about the positioning of the
   body in its environment. Brain processes are able to interpret
   this information.

4. {\underline{Body-World-Belief      Schema}}  The   body-world  schema has an
   extension that represents beliefs and other idealike structures.

5. {\underline {Records.}}  The B-W-B Schema are  representations that have the
   properties   required  for  records:  they  endure,  are  copiable,  and are
   combinable$^{11}$. These requirements  ensure that these representations are
   engraved in degrees of  freedom that can be  characterized as ``classical''.
   Superpositions  of such  classically  describable  states are  generally not
   classical. This  characterization of  ``classical'' (in terms of durability,
   copiability, and combinability) does not take one outside quantum theory: it
   merely distinguishes certain  functionally important kinds of quantum states
   from others.

6. {\underline{Evolution    Via the  Schoedinger   Equation.}} The  alert brain
   evolves under  the quantum  dynamical laws  from a state in  which one B-W-B
   schema is  excited to a state  in which a  quantum  superposition of several
   such states are  excited. That is,  the brain evolves  from a state in which
   one   B-W-B  schema  is   excited,  for a   period of   time   sufficient to
   ``facilitate'' the pattern, into a  quantum state that is a superposition of
   several  ``classical branches'', each  representing a  different classically
   describable state of the Body-World-Belief complex.

7. {\underline{The  Quantum Jump.}} The  Heisenberg actual  event occurs at the
   high-level of  brain activity  where the  different  classical branches have
   separated: this  event actualizes one  branch and  eradicates the others, in
   accord with  Heisenberg's idea of  what happens in a  measuring device.  The
   human brain is, in effect, treated as a quantum measuring device.

8. {\underline{Thoughts.}} The  occurrence of the Heisenberg event at this high
   level, rather than  at some lower  level (e.g., when  some individual neuron
   fires) is in line  with Wigner's  suggestion that the  reduction of the wave
   packet occurs in the  brain {\it{only}} at the  highest level of processing,
   where  conscious  thoughts  enter.  The  state of the  brain  collapses to a
   classical branch that encapsulates  and records the information contained in
   a classical description of the  body-world-belief complex.  It is postulated
   that this actualizing event at the level of the wave function  is associated
   with a conscious event that is a mental image of the information represented
   by the actualized B-W-B schema.

9. {\underline{Limitations.}   The theory  describes only  those collapses that
   occur in  the part  of the  physical world   associated with   human brains:
   Whether and where other events occur is left open. A parsimonious version of
   the theory  in which the  only  collapses are  those   associated with human
   brains  would account in  principle  for all  human  experience: there is no
   empirical   evidence  available today  that  would  demand any  other actual
   events. Such a parsimonious theory would be excessively anthropocentric. Yet
   any attempt to go beyond it would be  speculative in the absence of relevant
   data. In the parsimonious version  every actual event corresponds to a human
   thought, and every human thought  corresponds to an actual event: the theory
   is maximally linked to the empirical facts of human experience.

\newpage

\noindent{\bf{Appendix B. Survival Advantage} }

Contemporary  quantum  theory does not  have any  definite rule  that specifies
where the collapses occur. The  proposal adopted  here is designed to produce a
simultaneous resolution of the quantum  measurement problem and the mind-matter
problem.   Thus  the  proposal is  justified by   the fact  that it  produces a
coherent  model of  reality that  accords with our  actual  experience. Yet the
deeper question  arises: {\it Why}  should the world be  this way, and not some
other way?  {\it Why} should  the collapses be  to single  high-level classical
branches,   rather  than to  either   lower-level  states,  such as  firings of
individual neurons, or to higher-level  states that might include, for example,
many classical branches.

If we suppose that the determination  of where the collapses occur is fixed not
by some {\it a priori} principle  but by {\it habits}  that become ingrained
into nature,  or by some  yet-to-be-discovered  characteristic  of  matter that
does not single-out the  classical branches {\it ab  initio}, then the question
arises: Is the placement of the  collapses at high-level classical branches, as
specified in  our model,  favorable to  survival of  the  organism? If so, then
there would be an  evolutionary pressure for the  collapse location to migrate,
in our species, to this high-level placement.  The fact that the collapses, and
hence the accompanying experiences,  are classical and high-level would then be
consequences of underlying causes, rather than being simply an unexplained fact
of nature: it would be advantageous to  the survival of the  organism to
tie  whatever   fundamental  property   controls  collapses to  the  high-level
classical states of our model.

In fact, it is evident  that placement of the  collapses at a lower level would
introduce a disruptive stochastic element into the dynamical development of the
system. Any sort of dynamical process designed to allow the organism to respond
in an optimal way  to its  environmental situation would  have a tendency to be
disrupted by the introduction of stochastically instituted low-level collapses,
which will not  always be to  states that are  strictly  orthogonal. Thus there
would be an  evolutionary  pressure  that would  tend to push  the collapses to
higher levels. On  the other hand,  this pressure would  cease once the highest
possible level of classically specified branches is reached. The reason is that
in order for the organism  to {\it learn} there  must be records of what it has
done, and  these  records must  be able to  control  future  actions. But these
properties    are   essentially  the   properties  by  which  we  have  defined
``classical''.  Superpositions  of such  classical states  have, because of the
local  character of the  interaction terms in  the quantum  mechanical laws, no
ability  to   reproduce   themselves,  or to   control  future  actions  of the
organism.$^{11}$   Thus there  should be no  migration  of the  location of the
collapse to levels higher than those specified in our model.

\newpage

\noindent{\bf{Appendix C. Many-Worlds Theories.}}

I have accepted  here Heisenberg's  idea that there are  real events, that each
one represents a  transition from ``the possible''  to ``the actual'', and that
the  quantum   state can  be  regarded  as a    representation  of  ``objective
tendencies'' for such events to occur.  In fact, it is difficult to ascribe any
coherent meaning to the quantum state  in the absence of such events. For there
is then  nothing in the theory  for the  probabilities  represented by the wave
function to be probabilities {\it of~}: What does it mean to say that something
happens with probability $P$ if everything happens?

In our  model,  if we  say that  there is  no  collapse then  all the  branches
continue to  exist: there is  no singling out  and  actualization of one single
branch. Each of the  several branches will evolve  independently of the others,
and  hence it is   certainly  plausible to  say  that the  different  realms of
experience that we would  like to associate with  the different branches should
be independent  and non  communicating: the  records formed in  one branch will
control only that one branch, and have  no effect upon the others. But if there
is no  collapse then  it would  seem that  each of the   corresponding separate
branches should occur with probability  unity. Yet that would not yield a match
with  experience. In order to  get a match  with experience we  must be able to
effectively  discard in the  limit of an  infinite number of  repetitions of an
experiment those branches that have a quantum weight that tends to zero in this
limit. That is,  quantum states with  tiny quantum weights  should occur almost
never: they  should not  occur with  probability  unity! So  without some added
ontological  or  theoretical   structure the  many-worlds  (i.e.,  no-collapse)
theories  fail to give a  sensible  account of the  statistical  predictions of
quantum theory.

Of  course, the  key  question is  not whether  a  certain  experience X {\it
occurs}, but rather whether {\it my}  experience will be experience X. However,
the idea that many experiences occur, but that {\it my} experience will be only
one of  them  involves some  new sort  of  structure  involving  ``choice'' and
``my''. It  involves a  structure that goes beyond the  idea of a quantum
state of the world  evolving in  accordance with the  Schroedinger equation. At
that basic   quantum  level the  various  classically  describable branches are
components  that are  combined {\it   conjunctively}: the  universe consists of
branch 1 {\it and}   branch 2 {\it  and} branch 3 {\it  and} ... ; not branch 1
{\it or}  branch 2   {\it or}  branch 3 {\it  or} ... .  Yet the  world must be
decomposed  in terms of   {\it  alternative}  possibilities in  order to assign
different   statistical  weights to  the  different  components: the  {\it and}
composition  given by the   basic quantum  structure must be  converted into an
{\it or} composition. This   restructuring seems to require the introduction of
some new sort of beingness: the idea  of a psychological being that splits into
{\it  alternative} branches   while the  associated physical  body, evolving in
accord with the  Schroedinger  equation, is  splitting into a {\it conjunction}
of corresponding branches. By an  appropriate assignment of statistical weights
to the various alternative psychological branches one could then explain the
statistical   predictions  of  quantum  theory, but  this  would seem  to be an
ontological  {\it tour de  force}  compared  to  the simpler Wigner idea,
adopted here, that thoughts correspond to real Heisenberg-type events.

\newpage
\noindent{\bf References}

\begin{enumerate}
\item H.P. Stapp, {\it The Copenhagen Interpretation, } Amer. J. Phys.
{\bf{40}} 1098-1116 (1977) Reprinted in ref. 5.
\item J. von Neumann, {\it The Mathematematical Foundations of Quantum
Mechanics}, Princeton University Press (1955) (Translated from the original
(1932) German edition) Ch VI Sec. 1.
\item N. Weiner, {\it Back to Leibniz,} in Tech. Rev. {\bf 34} (9132), 201-203,
 222, 224; {\it Quantum Mechanics, Haldane, and Leibniz}, Philos. Sci.
{\bf 1} (1934), 479-482; {\it The Role of the Observer},
Philos. Sci.  {\bf 3} (1936), 307-319.
\item J.B.S. Haldane, {\it Quantum Mechanics as a Basis for Philosophy},
Philos. Sci. {\bf 1} (1934), 78-98.
\item H.P. Stapp, {\it Mind, Matter, and Quantum Mechanics,} Springer-Verlag
(1993).
\item S.M. Kosslyn, {\it Image and Brain}, MIT Press (1994).
\item E. Wigner, in {\it The Scientist Speculates}, ed. I.J. Good, Basic Books,
New York (1962).
\item A. Aspect, P. Grangier, and G. Roger, {\it Experimental Tests of Bell's
Inequalities using Time-varying Analysers}, Phys.
Rev. lett. {\bf 49} (1982), 1804-1807.
\item P.S. Churchland, {\it Neurophilosophy: Toward a Unified Theory of the
Mind/Brain}, MIT Press, Cambridge MA, 1992.
\item P.S. Churchland, ibid. , p.295.
\item H.P. Stapp, {\it Symposium on the foundations of modern physics 1990},
eds. P. Lahti and P. Mittelsteadt, World Scientific, Singapore, 1991.

\end{enumerate}

\end{document}